\documentstyle[12pt]{article}

\begin{document}
\setlength{\textheight}{1.2\textheight}


\def\beqn{\begin{equation}}
\def\eeqn{\end{equation}}
\def\beqa{\begin{eqnarray}}
\def\eeqa{\end{eqnarray}}

\def\ni{\noindent}
\def\dq{\delta q}
\def\R{{\cal R}}
\def\Im{\mbox{Im}}



\def\epaisfle{height1pt}
\def\longfle{width0.5cm}
\def\blancfle{\vskip1.8pt}
\def\espfle{\vskip0.2cm}
\def\espflc{\vskip1cm}

\def\flecher{\hbox{\vbox{\hrule \longfle \epaisfle
\blancfle}$\triangleright$}}

\def\flechel{\hbox{$\triangleleft$\vbox{\hrule \longfle \epaisfle
\blancfle}}}

\def\fleches{\vbox{\flecher \espfle \flechel \espflc}}


\def\epaisdir{height0.3pt}
\def\longdir{width0.4cm}
\def\blancdir{\vskip1.8pt}
\def\longdis{width2cm}

\def\direcr{\hbox{\vbox{\hrule \longdir \epaisdir
\blancdir}$\rangle$}}

\def\direcd{\hbox{$\langle$\vbox{\hrule \longdis \epaisdir
\blancdir}$\rangle$}}


\def\largmir{width0.2cm}
\def\epaishmir{height1pt}
\def\epaisvmir{width1pt}
\def\hautmir{height1.3cm}
\def\intmir{\hskip0.13cm}

\def\miroir{\vbox{\hrule \largmir \epaishmir
\hbox{\vrule \hautmir \epaisvmir \intmir \vrule \hautmir \epaisvmir}
\hrule \largmir \epaishmir}}


\def\esphforce{\hskip0.01cm}
\def\espvforce{\vskip0.9cm}
\def\esphrecul{\hskip0.01cm}
\def\esphmir{\hskip0.3cm}

\def\force{\vbox{\hbox{\esphforce \direcr} \hbox{$F+\delta
F$} \espvforce}}

\def\recul{\vbox{\hbox{\esphrecul \direcr} \hbox{$\delta q$}}}

\def\FD{
\hbox{\fleches \vbox{\hbox{\esphmir \miroir} \recul} \force}
}


\def\esphch{\hskip1.3cm}
\def\espvch{\vskip0.2cm}
\def\espvmir{\vskip0.85cm}
\def\esphq{\hskip0.9cm}
\def\esphd{\hskip0.1cm}

\def\champ{\hbox{\esphch \vbox{\flecher \espfle \flechel \espvch}}}

\def\imp{\vbox{\hbox{\esphforce \direcr} \hbox{$d p$}
\espvforce}}

\def\INTER{
\hbox{\vbox{\hbox{\esphd $\varphi$} \flechel \espvch \espvmir}
\vbox{\miroir \espvmir}\vbox{\hbox{\esphch $d N$} \blancdir \champ
\direcd \hbox{\esphq $q$}}\vbox{\miroir \espvmir}
\imp}
}


\def\espobl{\hskip0.14cm}
\def\espdev{\hskip0.2cm}
\def\chadeu{\vbox{\hbox{\espdev $u$} \flecher \espfle \flechel \espvch}}
\def\chadev{\vbox{\flecher \espfle \flechel \espvch}}
\def\esphrdev{\hskip1.7cm}

\def\boxit#1{\vbox{\hbox{\vbox{\hbox{\espobl}}#1}\hbox{$/$}}}
\def\vonde{\vbox{\hbox{$k$}}}

\def\onde{\boxit{\boxit{\boxit{\boxit{\vonde}}}}}

\def\DEVGEO{
\hbox{\vbox{\hbox{$x$} \espvmir}
\vbox{\miroir \espvmir}\vbox{\hbox{\espobl \chadev \onde \chadeu}
\hbox{\esphrdev \recul}} \vbox{\miroir \espvmir}}
}


\def\kp{K}
\def\kf{K_0}
\def\l{l}

\begin{titlepage}
\nopagebreak
\begin{center}
{\large\bf
QUANTUM LIMITS IN

SPACE-TIME MEASUREMENTS}

\vfill
        {\bf Marc-Thierry Jaekel\dag ~and Serge Reynaud\ddag }  \\
\end{center}
\dag Laboratoire de Physique Th\'eorique
 de l'Ecole Normale  Sup\'erieure\footnote{Unit\'e
propre du Centre National de la Recherche Scientifique, \\
associ\'ee \`a l'Ecole Normale Sup\'erieure et \`a l'Universit\'e
de Paris Sud.}(CNRS),
24 rue Lhomond, F75231 Paris Cedex 05, France \\
\ddag Laboratoire Kastler-Brossel\footnote{Unit\'e
de l'Ecole Normale Sup\'erieure et de l'Universit\'e Pierre et Marie
Curie,\\
 associ\'ee au Centre National de la Recherche
Scientifique.}(UPMC-ENS-CNRS), case 74,\\
 4 place Jussieu,  F75252 Paris Cedex 05,
France \\

\vfill

\vfill

\begin{abstract}

Quantum fluctuations impose fundamental limits on measurement
and space-time probing. Although using optimised probe fields
can allow to push sensitivity in a position measurement
 beyond the "standard quantum limit", quantum fluctuations of the
probe field still result in limitations which are determined by
irreducible dissipation mechanisms. Fluctuation-dissipation relations
in vacuum characterise the mechanical effects of radiation pressure vacuum
fluctuations, which lead to an ultimate quantum noise for
positions.
For macroscopic reflectors, the quantum noise
on positions is dominated by gravitational vacuum fluctuations,
and takes a universal form deduced from
quantum fluctuations of space-time curvatures in vacuum.
These can be considered as ultimate space-time fluctuations,
fixing ultimate quantum limits in space-time measurements.

\end{abstract}
\vfill

\begin{flushleft}

\vfill
        {\normalsize LPTENS 95/13 \\
May 1995\\
}

\end{flushleft}

\end{titlepage}

{\bf I. Introduction}

\bigskip
Interest in high sensitivity measurements of position has been
recently revived, under the impulse of projects for
detecting gravitational waves \cite{Virgo}.
The required very high level of sensitivity
has led to consider the role of limitations due to
quantum fluctuations \cite{Braginsky}.
Limits imposed by quantum mechanics give rise to
practical and theoretical problems. When disregarding noise sources
(like seismic or thermal noises),
which although dominant can be minimised in principle,
fundamental limitations subsist which cannot be bypassed.

An early argument related to Heisenberg's microscope
asserts
that measurement of a quantity $q$ with a precision (mean
square deviation) $\Delta q$ must
induce a perturbation of its conjugate variable $p$ of an amount
$\Delta p$ such that ($\hbar$ is Planck constant):
\beqn
\label{hi}
\Delta q \Delta p \ge {\hbar \over 2}
\eeqn
Non-commutativity of quantum variables then puts limits on
successive measurements, and impedes a continuous
and independent determination of positions in space-time. For instance,
successive independent measurements of position $q(t)$ and
$q(t+\tau)$ of a free mass $m$ are subject to
a "standard quantum limit" \cite{Caves}:
$$\Delta q(t+\tau)^2 + \Delta q(t)^2 \ge  2\Delta q(t+\tau) \Delta q(t)$$
$$\ge {\hbar\over m}\tau =  \Delta q^2_{SQL}$$
As extensively discussed in last years,
successive independent measurements may not correspond to
the best measurement strategy \cite{YCO,BK}.
As has been shown, when optimising the
measurement strategy, the "standard quantum limit" can be beaten
leading to actual limitations which are much lower \cite{JR1,PCW}.
Here, we discuss those ultimate limitations which subsist as a
consequence of the quantum noise induced by quantum field
fluctuations in vacuum.

\bigskip
\bigskip
{\bf II. Probe field quantum fluctuations}

\bigskip
We first briefly recall the quantum limits imposed by quantum
fluctuations of the probe field in a measurement of
the position of a mirror using interferometry.
An interferometric measurement of position can be seen as
a phase measurement for the probe field. On one output
port of an interferometer, a signal is detected
which varies like the difference of phase-shifts
$\varphi$ undergone by the probe field
in different arms. For a
monochromatic plane wave with wave-vector $\kf$,
phase-shift variations directly
provide an estimator for variations of one of the mirrors' position
$q$:
$$\varphi(t) = 2 \kf q(t)$$

\bigskip
\hskip4.6cm \INTER
\begin{center}
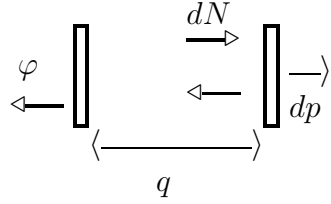
Back action during measurement with a probe field

$Figure \quad 1$
\end{center}

Phase fluctuations of the probe field directly affect a position
measurement. If the probe field intensity is increased,
in order to improve signal-to-noise ratio by increasing the number of
detected photons, then another source of noise, related to
intensity fluctuations of the probe field ($I = dN / dt$, the time
derivative of photon number $N$)
also increases.
As the probe field exerts a radiation pressure on the
mirrors, intensity fluctuations affect the momentum $p$ of the measured mirror
and consequently its position (see Figure 1):
$$F = {d p\over d t}
= 2 \hbar \kf {d N\over d t}
= 2 \hbar \kf I$$
Coupled field and mirror provide an example of back action during
measurement.
The measured phase $\varphi$ is also affected
by fluctuations of its conjugate variable $N$.
As fluctuations of conjugate variables are constrained by Heisenberg
type inequalities, phase and intensity fluctuations of the probe
field finally put limits on the allowed sensitivity in an
interferometric measurement of position. When using coherent light as
input fields, an optimum is reached when phase and intensity
fluctuations provide equally important sources of noise, leading to
the "standard quantum limit". Such limit relies
on the independent character of the two conjugate sources of noise,
which are related to the particular input fields used.

To discuss the general case, it is sufficient to analyse the
effects of probe field fluctuations on measurement sensitivity
in a linearised treatment of fluctuations \cite{JR1}.
During measurement, the probe field adds a noise $q_n$ which is best
described by its Fourier components:
$$q_n(t) = \int_{-\infty}^\infty {d\omega \over 2\pi}
e^{-i\omega t} q_n[\omega]$$
and which is the sum of a noise related to the phase of the probe
field and of a noise related to the probe field intensity:
\beqn
\label{qn}
q_n =  {\delta \varphi \over 2 \kf} \; + \;
\chi_{qq} \; 2\hbar \kf \delta I
\eeqn
Noise frequencies are in mechanical range and are typically much
smaller than probe field optical frequencies, so that phase and
intensity variations appear as modulations of two conjugate
quadrature components of the input fields. Variations of position
due to intensity fluctuations can be treated in linear response
formalism, so that
$\chi_{qq}$ describes the mirror's response to an applied force:
$$\delta q [\omega] = \chi_{qq}[\omega] \delta F[\omega]$$
For simplicity, one can consider that the mirror, of mass $m$,
is mechanically bound
with a proper frequency $\omega_0$, and that all dissipative
couplings can be summarised in a friction coefficient $\gamma$
depending on the frequency:
$$\chi_{qq}[\omega] = {1 \over m (\omega_0^2 -\omega^2 - i \gamma \omega)}$$

Correlation functions of the noise added in the measurement
 are then characterised by their spectrum:
$$<q_n(t)q_n(0)> -<q_n(t)><q_n(0)> =  C_{q_nq_n}(t)$$
$$ C_{q_nq_n}(t) = \int_{-\infty}^\infty {d\omega \over 2\pi}
e^{-i\omega t} C_{q_nq_n}[\omega]$$
Constraints on spectra
describing phase and intensity fluctuations, and
correlations between phase and intensity, are put by
Heisenberg type inequalities. They then determine lower bounds for
spectra of the noise variable $q_n$. Several cases, corresponding to
different measurement strategies must be distinguished.

For an antenna, like a
gravitational wave detector, the limit fixed on sensitivity is
best expressed by the minimal spectral energy density of
the remaining noise.
For probe fields with uncorrelated phase and intensity,
Heisenberg inequalities for probe field variables take the usual
form, similar to (\ref{hi}), for each frequency, and
lead to the "standard quantum limit":

For a free mirror ($\omega_0 \simeq 0$):
\beqn
\label{sql}
{1\over2}m\omega^2 C_{q_n q_n}[\omega] \ge
\hbar m\omega^2 |\chi_{qq}[\omega]|
\simeq \hbar
\eeqn
$$\Delta q_n^2 = C_{q_nq_n}(0) \simeq \Delta q_{SQL}^2 = {\hbar \over m} \tau
 \qquad \qquad
\tau \simeq {\Delta \omega \over \omega^2}$$
($\Delta \omega$ is the detection bandwidth, providing a time
parameter $\tau$ characteristic of measurement).
The "standard quantum limit" is given by the modulus of the mirror's
mechanical response function, i.e. essentially by its reactive part,
and corresponds to a constant spectral energy density
equal to $\hbar$.

In an optimal measurement, correlations of
phase and intensity of the probe field must be adapted to
the mechanical response function of the mirror. In other words,
input fields must be used which minimise fluctuations for the
particular combination of the two quadrature components which finally
enters the output noise $q_n$ (see (\ref{qn})),
at the expense of increasing fluctuations
for the conjugate combination. Squeezing of input fields
must also be realised in different directions for different frequencies, as
specified by the mechanical response function of the mirror.
When considering such ideally prepared input fields,
measurement sensitivity is then only limited by
the dissipative part of the mirror's mechanical response:
\beqn
\label{uql}
{1\over2}m\omega^2 C_{q_n q_n}[\omega] \ge
 \hbar m\omega^2 |\Im \chi_{qq}[\omega]|
\simeq \hbar {\gamma \over \omega}
\eeqn
Such limit is much lower than
the "standard quantum limit" (\ref{sql}).

For optimized measurement strategy, quantum fluctuations still put a limit on
sensitivity, which
is determined by dissipation mechanisms \cite{BK,JR1}.
As a result of the
coupling of the mirror to the probe field,
that is to a system with infinitely
many degrees of freedom, the mirror's dynamics necessarily contain a
minimal dissipative component related to quantum field fluctuations.
 This dissipative part again contains an irreducible contribution
due to vacuum field fluctuations, which can then be seen as fixing
an ultimate limitation
on measurement sensitivity.

\bigskip
\bigskip
{\bf III. Quantum fluctuations of position in vacuum}

\bigskip
In this part, we discuss the dissipative contribution to the
mirror's motion due to
vacuum field fluctuations, i.e. in a state with no photons, and the resulting
fluctuation-dissipation relations for the position of a mirror in
vacuum. Quite generally,
quantum field fluctuations induce quantum fluctuations of field stress-tensors,
so that a mirror is submitted to a fluctuating radiation pressure (or
a fluctuating force $F$)
due to quantum fluctuations of the incident field:
$$<F(t) F(0)> - <F>^2 =
 \int_{-\infty}^\infty {d\omega \over 2\pi} e^{-i\omega t}
C_{FF}[\omega]$$
As a result of general principles which govern motion in a fluctuating
environment \cite{Einstein}, the mirror is also submitted to
an additional force when it moves. The motional force can be described within
linear response formalism \cite{Kubo}, by a motional susceptibility
$\chi_{FF}$:
$$\delta F[\omega] = \chi_{FF}[\omega] \dq[\omega]$$
which is related to the generator of the perturbation
(that is $F$ for a displacement), and
which satisfies a fluctuation-dissipation relation (see Figure 2):
\beqn
\label{fd}
2 \hbar \Im \chi_{FF}[\omega] = C_{FF}[\omega] - C_{FF}[-\omega]
\eeqn

\newpage
\hskip5.5cm \FD
\begin{center}
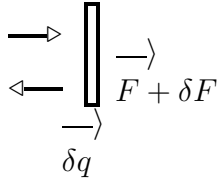
Fluctuating force at rest and force induced by motion

$Figure \quad 2$
\end{center}

The mechanical response of the mirror to an applied force
is then modified by its coupling to the probe field,
 and necessarily
contains a dissipative part related to fluctuations of the
incident field radiation pressure:
\beqn
\label{mrf}
\chi_{qq}[\omega] = {1 \over m (\omega_0^2 - \omega^2) - \chi_{FF}[\omega]}
\eeqn
The dissipative contribution due to quantum field fluctuations depends on
 the input field state (\ref{fd}), but always includes a
part which cannot be eliminated, and which is due to vacuum field fluctuations.
Vacuum can also be seen as the state of thermodynamic equilibrium
at zero temperature, so that it satisfies a further fluctuation-dissipation
relation which completely determines fluctuation spectra from commutators only:
$$C_{FF}[\omega] = 2 \hbar \theta(\omega) \Im \chi_{FF}[\omega]$$
$\theta(\omega)$ is Heaviside step function, ensuring that no excitations
with negative energy can exist in vacuum.
Vacuum quantum field fluctuations impose
an irreducible dissipative contribution \cite{JR3}:
\beqn
\label{vfd}
\chi_{FF}[\omega] = i m \gamma \omega =
i \alpha {\hbar \omega^3 \over c^2}
\eeqn
where $c$ is light velocity, and $\alpha$ is a dimensionless factor
depending on the geometry and particular coupling between mirror and field
(Lorentz invariance of vacuum also implies
that the motional force does not contain
any contribution proportional to the velocity).

Further fluctuation-dissipation relations characterise
motion in vacuum.
Under the effect of force fluctuations in vacuum, the mirror is submitted to
a quantum Brownian motion which results in
fluctuations of its position:
$$<q(t) q(0)> - <q>^2 =
 \int_{-\infty}^\infty {d\omega \over 2\pi} e^{-i\omega t}
C_{qq}[\omega]$$
The coupled system consisting of the mirror's position and field radiation
pressure can be treated consistently within linear response formalism,
and shown to
also satisfy fluctuation-dissipation relations \cite{JR6}. In particular, the
position of a mirror coupled to vacuum fields
satisfies both relations, that typical of
linear response to an external perturbation:
$$2 \hbar \Im \chi_{qq}[\omega] = C_{qq}[\omega] - C_{qq}[-\omega]$$
and also that typical of the zero-temperature limit of thermal equilibrium:
$$ C_{qq}[\omega] = 2 \hbar \theta(\omega) \Im \chi_{qq}[\omega]$$
In vacuum, positions have quantum fluctuations with spectra which
are determined by the dissipative part of the mirror's mechanical admittance
((\ref{mrf}) and (\ref{vfd})).
Outside proper resonance frequencies, fluctuations are those induced by
radiation pressure of vacuum fields.
For a free mirror ($\omega_0 \simeq 0$):
\beqn
\label{uqn}
{1\over2} m \omega^2 C_{qq}[\omega] \simeq \hbar \theta(\omega)
{\gamma \over \omega}
\eeqn
$${\gamma \over \omega} = \alpha {\hbar\omega \over m c^2} \ll 1$$
Coming back to the discussion of a position measurement, quantum
fluctuations of the mirror's position include permanent fluctuations
induced by radiation pressure of vacuum fields (due to field
fluctuations with frequencies comprised between $0$ and $\omega$),
and added fluctuations due to probe field fluctuations (field
frequencies around the probe frequency $c\kf$).
Taking these two noises into account again leads to a quantum limit
for the sensitivity in an optimal measurement which is fixed by the
dissipative part of the mirror's mechanical response (\ref{uql})
\cite{JR6}. Hence,
vacuum field fluctuations induce an ultimate quantum noise
on position.
Its order of magnitude is determined by the Compton wave-length
corresponding to the reflector's mass.  For macroscopic mirrors,
(of mass greater than Planck mass $\sim 22 \mu\mbox{g}$),
the Compton wave-length becomes smaller than Planck length $\sim 1.6
\; 10^{-35}$m, so that
this limit is actually negligeable. At such level, quantum fluctuations
related to gravitation must be taken into account in order to discuss
actual quantum limits.

\bigskip
\bigskip
{\bf IV. Gravitational quantum fluctuations}

\bigskip
Quantum fluctuations of gravitation must limit the determination of positions
with a precision at the level of Planck length \cite{Wheeler}. Indeed, quantum
fluctuations of space-time metric affect length measurements.
This perturbation can be described intrinsically,
that is independently of a particular choice of a reference system,
in terms of space-time curvatures.
When propagating in space-time,
a probe field registers curvature fluctuations. The main effect of
metric perturbations with small wave-vectors $k$ (when compared to
probe field wave-vector $\kp$) is obtained in the eikonal approximation.
Then, the probe field momentum, or wave-vector $\kp_\mu = \kp_0 u_\mu$,
follows the law of geodesic deviation \cite{BM}:
$$\partial_\nu \kp_\lambda = \int_0^\l R_{\lambda\mu\nu\rho}(x - u \sigma)
\kp^\mu u^\rho d\sigma$$
Variations of momentum integrate curvature perturbations, described
by their Riemann tensor $R_{\lambda\mu\nu\rho} $, encountered
during propagation (along direction $u$, $u^0 = 1$) from the
emitter to the receptor of coordinates $x$ ($\l$ is the total propagation
length and $\sigma$ an affine parameter).
In particular, phase shifts of the probe field
can be obtained from frequency deviations ($c \kp_0$ being the time
derivative of the phase). The corresponding expression coincides
with the formula giving the red-shift induced by a stochastic
background of gravitational
waves \cite{MG}.
It can also be written
as a variation of the measured distance $q$ (depending on time $t =
x^0 / c$) due to curvature
fluctuations (of Fourier components $R_{\lambda\mu\nu\rho}[k]$) (see
Figure 3):
\beqn
\label{gd}
{1 \over c^2}{d^2 \dq\over dt^2} = \int_0^\l d\sigma \int {dk \over (2\pi)^4}
e^{-ik.(x-u\sigma)} R_{0\mu0\rho}[k]u^\mu u^\rho
\eeqn

\bigskip
\hskip5cm \DEVGEO
\begin{center}
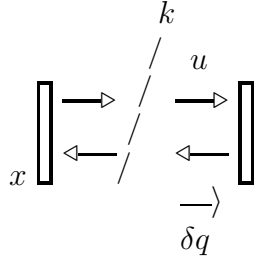
Gravitational perturbation of probe field propagation

$Figure \quad 3$
\end{center}

\bigskip
Quantum fluctuations of the metric can be treated like those of other
elementary fields \cite{FW}. Linearised Einstein equations provide
the graviton propagator, and hence, in agreement with
fluctuation-dissipation relations, metric quantum fluctuations
in vacuum \cite{UdW,JR11}. These are
characterised by Planck length ($G$ is Newton's gravitation
constant):
$$l_p = ({\hbar G \over c^3})^{1\over2} \sim 1.6 \; 10^{-35} \mbox{m}$$
At lowest order in $l_p$, metric vacuum fluctuations
are determined by gravitational waves zero-point fluctuations.
Corresponding space-time curvature fluctuations provide vacuum
fluctuations which are invariant under gauge symmetries (i.e. metric
transformations
taking the form of those induced by changes of coordinates):
$$C_{R_{\lambda\mu\rho\nu}R_{\lambda'\mu'\rho'\nu'}} [k] =
16\pi^2 l_p^2\theta(k_0)\delta(k^2) \times$$
$$\times (\R_{\lambda\mu\lambda'\mu'} \R_{\rho\nu\rho'\nu'}
+ \R_{\lambda\mu\rho'\nu'} \R_{\rho\nu\lambda'\mu'}
- \R_{\lambda\mu\rho\nu} \R_{\lambda'\mu'\rho'\nu'})$$
$$\R_{\lambda\mu\rho\nu} = {1\over2}(k_\lambda k_\rho \eta_{\mu\nu}
+ k_\mu k_\nu \eta_{\lambda\rho} - k_\mu k_\rho \eta_{\lambda\nu}
- k_\lambda k_\nu \eta_{\mu\rho})$$
Riemann curvature fluctuations can also be determined from
Lorentz and gauge invariance in vacuum, symmetry properties of
Riemann tensor, correlations of Einstein tensor (which vanish) and
normalisation to ${1\over2}\hbar$ of spectral energy density.

When integrated along propagation of the probe field (see
(\ref{gd})),
curvature fluctuations induce
fluctuations of distances \cite{JR10}:
$$C_{qq}[\omega] = \beta l_p^2 {\theta(\omega) \over \omega}$$
where $\beta$ is a geometric factor depending on the particular
measurement technique used
(one-way or round trip probing).
One should note that
for low frequencies, that is for
frequencies well below Planck frequency,
spectra of distance fluctuations induced by gravitational
fluctuations only depend on the assumed effective behavior of gravitation
at low frequencies (as described by Einstein theory)
and conformity of vacuum fluctuations with fluctuation-dissipation
relations.
As a consequence of characteristic properties in vacuum,
spectra of distance fluctuations only contain positive frequencies,
so that when rewritten in space-time domain,
distance correlations are not symmetric under exchange of
their arguments.
Geodesic distances then appear as non-commutative quantum variables.

For low frequencies, distance fluctuations induced by gravitational
fluctuations take a form which is similar to that of fluctuations induced by
radiation pressure. They however differ in their order of magnitude,
gravitational fluctuations imposing a limit on
space-time probing of the order of Planck length.
For an optimal measurement, there results two regimes
of ultimate space-time fluctuations,
depending on the endpoint mass used.

For a "microscopic" mass, i.e. smaller than Planck mass:
$$m \ll m_p \qquad \qquad
m_p = ({\hbar c \over G})^{1\over2} \simeq 22 \mu \mbox{g}$$
radiation pressure fluctuations dominate,
and the noise spectrum depends on the mass (see ({\ref{uqn})):
$$C_{qq}[\omega] \simeq \lambda_c^2  {\theta(\omega) \over \omega}
\qquad \qquad \lambda_c = {\hbar \over m c}$$
For a "macroscopic" mass, i.e. greater than Planck mass:
$$m \gg m_p$$
metric fluctuations dominate and
the noise spectrum is universal:
$$ C_{qq}[\omega] \simeq l_p^2 {\theta(\omega) \over \omega}$$

Optimal space-time probing is thus obtained using "macroscopic"
end-point reflectors. Measurement of positions in space-time is then
limited by an ultimate quantum noise which is universal, as it only
depends on universal constants through Planck mass.
Quantum fluctuations of curvature in vacuum can thus be considered
as giving rise to
ultimate space-time fluctuations \cite{JR10}.

\bigskip
\bigskip
{\bf V. Conclusion}

\bigskip
When optimising the input probe fields used,
sensitivity in an interferometric measurement of position
can be pushed beyond the "standard quantum limit", and becomes
limited by dissipative mechanisms only. Vacuum field fluctuations induce
quantum fluctuations of radiation pressure which fix an
irreducible dissipative part for the mechanical response function of
the measured mirror. As a result of fluctuation-dissipation
relations, sensitivity in a position measurement
can be seen to be limited by
an ultimate quantum noise due to quantum field fluctuations in
vacuum. For reflectors with a mass greater than Planck mass,
 the quantum noise induced by
vacuum radiation pressure fluctuations becomes negligeable, and
position fluctuations are dominated by metric vacuum fluctuations,
which are universal and have an order of magnitude of Planck length.
Optimal measurements in space-time
are thus ultimately limited by quantum fluctuations of space-time
itself, due to quantum fluctuations of space-time curvatures in
vacuum. The non-commutative nature of space-time geometry
already reveals itself at low frequencies, in the ultimate quantum limits
 imposed to space-time measurements.

\end{document}